\documentclass[doublecol]{epl2} 
\usepackage[margin=1.85cm]{geometry}
\usepackage{amsmath}
\usepackage{amsmath}
\usepackage{amsfonts}
\usepackage{amssymb}
\usepackage{graphicx}
\usepackage{mathabx}
\usepackage{lmodern}
\usepackage{cuted}
\usepackage{hyperref}
\usepackage{dcolumn}
\usepackage{bm}
\usepackage[mathlines]{lineno}
\usepackage{float}
\usepackage[T1]{fontenc}
\usepackage{subfigure}
\title{A Realistic Neutrino mixing scheme arising from $A_4$ symmetry.}
\author{Manash Dey\inst{1}\footnote{manashdey@gauhati.ac.in} \and Subhankar Roy\inst{1}\footnote{subhankar@gauhati.ac.in\,(Corresponding author)}}
\shortauthor{\small{M. Dey and S. Roy}}
\institute{                    
  \inst{1} Department of Physics, Gauhati University - Gopinath Bordoloi Nagar, Jalukbari, India, 781014\\
}
\abstract{
We propose a unique lepton mixing scheme and its association with an exact hierarchy-philic neutrino mass matrix texture in the light of Type-I+Type-II seesaw mechanism under the framework of $A_4 \times  Z_{10}$ discrete flavour symmetry. The proposed model successfully predicts the normal ordering of neutrino masses and the two Majorana phases. Additionally, the analysis extends to the effective Majorana neutrino mass, in the context of neutrinoless double beta\,($0\nu\beta\beta$)-decay.}
\begin{document}
\maketitle
\section{Introduction}

It is believed that the answers to the greatest mysteries of the universe are related to some tiniest particles, and neutrino is one of them. After the discovery of neutrinos in 1950s \cite{Cowan:1956rrn}, these particles have constantly surprised us. Neutrinos have established themselves as an essential part of the Standard Model\,(SM) of elementary particle physics. In the light of the SM, neutrinos are massless particles. But, in the recent past, results from various experiments \cite{ SNO:2002tuh, KamLAND:2002uet, Super-Kamiokande:1998uiq} have successfully verified the phenomenon of Neutrino Oscillation which is a quantum mechanical theory. Neutrino oscillation was first put forward by Bruno Pontecorvo in the year 1957 \cite{Pontecorvo:1957cp}. It shares the fact that the three neutrino mass eigenstates\,($\nu_{i= 1,2,3}$) mix up to produce the three flavour neutrino states\,($\nu_{l= e,\mu,\tau}$). Experimentally it is observed that the flavour states can convert themselves from one to the other, and this hints at the existence of non-zero and non-degenerate tiny neutrino masses. In this connection, the SM has two primary shortcomings: it cannot fully explain why neutrinos of different flavours mix together in a peculiar way, and it struggles to account for the extremely small masses of neutrinos. The search for answers to these questions has taken us to the domain of neutrino mass textures, which is a phenomenological concept Beyond the Standard Model\,(BSM). By the word ``neutrino mass texture'' we understand certain correlations or constraints over the neutrino mass matrix elements. In this light, several ideas such as texture zeroes\,\cite{Ramond:1993kv, Ibarra:2003Jb}, $\mu$-$\tau$ symmetry\,\cite{Harrison:2002er}, $\mu$-$\tau$ mixed symmetry\,\cite{Dey:2022qpu}, $\mu$-$\tau$ reflection symmetry\,\cite{Xing:2022uax}, $\mu$-$\tau$ antisymmetry\,\cite{Xing:2015fdg, Xing:2020ijf} etc. are attributed to the neutrino mass matrix $M_{\nu}$ with an aim to find relationships among the neutrino mixing parameters. The neutrino mass matrix ($M_{\nu}$) carries the information of all the observational parameters namely, solar mixing angle\,($\theta_{12}$), reactor mixing angle\,($\theta_{13}$), atmospheric mixing angle\,($\theta_{23}$), three mass eigenvalues \, $m_1, m_2$ and $m_3$ and a Dirac CP violating phase\,($\delta$). In addition to these, $M_{\nu}$ shelters two Majorana phases $\alpha$ and $\beta$. From the neutrino oscillation experiments, we can estimate the mixing angles and $\delta$. However, the neutrino oscillation experiments cannot give the exact values of the neutrino mass eigenvalues and the two Majorana phases. Hence, the investigation of the phenomenological ideas based on textures of the neutrino mass matrix\,($M_{\nu}$) is still relevant. However, to probe extremely small neutrino masses, these textures are generally formulated within the framework of different neutrino mass generation mechanisms\,\cite{Ohlsson:2013xva, King:2003jb}, and among the numerous mechanisms explored, Seesaw has emerged to be the widely utilized one. In our present work, we emphasize on seesaw mechanisms viz., Type-I\,\cite{Cai:2017mow, Mohapatra:2004zh, King:2003jb} and Type-II\,\cite{King:2003jb, Cheng:1980qt}. The Type-I seesaw mechanism, is a simple extension of the SM, in which right handed neutrinos, $\nu_R$ are supplemented in order to generate Majorana and Dirac mass terms, and they are generally of the forms $\sim M_R (\bar{\nu^c_R}\,\nu_{R})$ and $\sim M_D (\bar{\nu}_L \nu_R)$ respectively. Here $\nu_L$ is the left-handed neutrino field of the SM lepton doublet $D_{l_L}$; $M_R$ and $M_D$ are complex square matrices which are in general non-diagonal. If we approximate that the scale of $M_R$ is greater than that of $M_D$, then the neutrino mass matrix under Type-I seesaw is given by,
\begin{equation}
-M_D \cdot M^{-1}_R \cdot M^T_D,
\end{equation}
which are found to be of the order $\sim 10^{-2}\,eV$. Therefore, the smallness of neutrino masses in this case is attributed to the heavy scale of $M_R$, which acts like a seesaw. The Type-II seesaw mechanism, on the other hand, is an extension of the higgs sector of the SM with a heavy $SU(2)_L$ triplet\,($\Delta$). The Majorana neutrino mass term, in this scenario, appears as $\sim y(\bar{D}_{l_L}\, D_{l_L}^c)\,{i \sigma_2 \Delta} $. In the limit where $m_{\Delta} \gg v_{EW}$, the neutral component of the $\Delta$ triplet field acquires a non zero vacuum expectation value\,(VEV) to account for small Majorana neutrino mass of the form $\sim y\, v_{\Delta}$, where $v_{\Delta}\, \propto v^2_{EW}/\text{m}^2_{\Delta}$ is the VEV of the neutral component of the higgs triplet\,($\Delta^{\circ}$), $\text{m}_{\Delta}$ is the mass of the triplet field\,($\Delta$), $v_{EW}$ is the SM Higgs VEV, and $y$ is the corresponding coupling constant. The Type-I and Type-II seesaw mechanisms are often fused together\,\cite{Wong:2022qyg, Akhmedov:2006de}, and the latter is coined as Hybrid seesaw mechanism in the literature\,\cite{Cai:2017mow, Akhmedov:2006de, Wong:2022qyg}. By incorporating this mechanism, one can build effective Majorana neutrino mass matrices ($M_{\nu}$) and can explore the possibilities of new lepton mixing patterns. 
The information of all the three neutrino masses, three mixing angles and the three CP-violating phases are contained within the neutrino mass matrix. In that way, the neutrino mass matrix is an important element that reflects all aspects of neutrino oscillation and beyond. Hence, while designing a model on leptons, the texture or pattern of the neutrino mass matrix is to be emphasized. For example, how many parameters it contains, or what are the relevant independent parameters like Yukawa couplings and vacuum expectation values appearing from the model are to be taken into consideration.  We define the independent quantities like yukawa couplings and VEVs as the \emph{model parameters}. Now in principle, depending on the chosen model, the number of model parameters could be unconstrained. It may so happen that they combine (not necessarily in a linear way) and generate some specific elements within the mass matrix. We define the latter as \emph{texture} parameters. We see that the number of texture parameters is constrained and cannot exceed six.  It may so happen, that the individual model parameters are not derivable, whereas the the texture parameters are. We emphasize the independence of the texture parameters. Now, if the mass matrix contains less than six texture parameters, then it implies that with lesser number of independent parameters we can predict more physical ones which happens to be one of the major goals of the model builders. One may derive a model independent phenomenologically neutrino mass matrix with defined number of independent texture parameters. However, it becomes challenging to derive the same texture following a top-down approach.	
	The plan of the paper is given as follows: In the next section, we discuss the neutrino mixing matrix and how our parametrisation is unique as well as helpful. In the third section, we talk about the model used to build the effective Majorana neutrino mass matrix texture. The numerical analysis and important observations are highlighted in the fourth section. Finally, in the fifth section, we write the summary and discussion of the present work.
\section{Parametrization of the PMNS matrix}\label{section2}
Apart from $M_{\nu}$, the neutrino physics phenomenology relies on another important quantity $U$, which is widely known as the Pontecorvo-Maki-Nakagawa-Sakata matrix\,
(PMNS matrix)\,\cite{Maki:1962mu} or the neutrino mixing matrix. The mixing matrix $U$ is a $3 \times 3$ unitary matrix and contains 9 free parameters. We demarcate $U$ which is the general one, from the $\tilde{U}$ that excludes all phases except $\delta$. These two mixing matrices are related to each other through the following transformation:
\begin{equation}
U= P_{\phi} \cdot \tilde{U} \cdot P_M ,
\end{equation}
a detailed discussion can be found in Appendix\,(see eq.\ref{pmns}). In our forthcoming discussions, we intend to employ the mixing matrix $\tilde{U}$.
In the realm of particle physics, several mixing schemes are proposed; such as Bimaximal (BM) mixing\cite{Barger:1998ta}, Tri-bimaximal (TB) mixing\cite{Harrison:2002er}, Hexagonal (HEX) mixing\cite{Albright:2010ap}, and  Golden ratio (GRa) mixing\cite{Ding:2011cm}. These schemes were closely examined in light of experimental data available at the time, revealing intriguing connections with a specific texture known as $\mu$ -$\tau$ symmetry\,\cite{Harrison:2002er}. Within the framework of  $\mu$-$\tau$ symmetry, the mixing angle $\theta_{12}$ is set free, while $\theta_{23}$ and $\theta_{13}$ are constrained to $\sin^{-1}(\frac{1}{\sqrt{2}})$ and 0, respectively. Notably, the selection of these angles appears to be special. For instance, when $\theta_{12}$ is set to $\sin^{-1}(\frac{1}{\sqrt{3}})$ (TB mixing), $\mu$ -$\tau$ symmetry naturally emerges through associated discrete symmetries like $A_4, S_4$ and $\Delta(54)$. Similarly, if $\theta_{12}$ is assigned $\sin^{-1}(1/\sqrt{1+\phi^2})$\,(GR mixing), where $\phi=(1+\sqrt{5})/2$ is the golden ratio, then the $\mu$ -$\tau$ symmetric mass matrix naturally emerges from $A_5$ symmetry. What may be concluded is that the choices of the mixing angles are indeed very special in the sense of their association with particular textures, and, more importantly, with certain geometries or symmetries. After these choices were put forward, the model-building community responded strongly, and this led to the investigation of various discrete flavour symmetries in the realm of neutrino physics phenomenology. A certain mixing scheme which is consistent with the experiments as well as with specific model based on discrete flavour symmetry, is indeed very special and, we realise that the TBM, BM,  HEX and GR mixings are the perfect examples highlighting this fact. However, the recent advancements in the neutrino physics experiments\,\cite{Gonzalez-Garcia:2021dve} have resulted in a non-zero $\theta_{13}$ and hence the $\mu$-$\tau$ symmetry and the associated mixing schemes cannot be accepted without correction. On the other hand, one may opt for an intelligent choice of a fresh lepton mixing matrix and further try to understand the framework on which it's grounded. With this motivation, we posit a lepton mixing scheme as
\begin{gather*} 
\sin{\theta_{12}}= 1/\sqrt{3},\quad \sin{\theta_{13}}= 1/(4\sqrt{3}),\quad\sin{\theta_{23}}= \sqrt{5}/3,\\\delta = 270^{\circ},\nonumber
\end{gather*}
suggesting a $\tilde{U}$ consistent with the experiments\,\cite{Gonzalez-Garcia:2021dve} which is shown as in the following,
\begin{equation}
\label{PMNS1}
 \tilde{U} =
 \begin{bmatrix}
 \frac{1}{6}\sqrt{\frac{47}{2}} & \frac{\sqrt{47}}{12} & \frac{i}{4\sqrt{3}}\\
 \frac{1}{18}\,i\,\sqrt{\frac{5}{2}}-\frac{2}{3 \sqrt{3}} & \frac{2}{3}\sqrt{\frac{2}{3}}+ i \frac{\sqrt{5}}{36} & \frac{1}{12}\sqrt{\frac{235}{3}}\\
 \frac{1}{3}\sqrt{\frac{5}{3}}+ \frac{i}{9 \sqrt{2}}& \frac{i}{18}-\frac{1}{3}\sqrt{\frac{10}{3}}  & \frac{1}{6}\sqrt{\frac{47}{3}} \\
 \end{bmatrix}.
\end{equation}
In the upcoming section, we shall explore how this mixing scheme can be realised in the light of discrete flavour symmetry.
\section{The Model}
\label{section3}
We extend the standard model group with $A_{4}$ symmetry\,\cite{Ma:2001dn} in the framework of a Type-I+Type-II seesaw mechanism. We enhance the field content with three heavy right handed singlet neutrino fields ($\nu_{e_R}$, $\nu_{\mu_R}$, $\nu_{\tau_R}$) which transform as ($1$, $1^{''}$, $1^{'}$) under $A_{4}$. Also, we consider three $A_{4}$ scalar triplets $\phi$, $\Phi$ and $\Delta$, which individually transform as $2$, $2$ and $3$ respectively under $SU(2)_L$. In addition, four scalar singlets $\xi$, $\zeta$, $\eta$ and $\kappa$ are introduced which transform as $1$, $1$, $1^{''}$ and $1^{'}$ respectively under $A_4$. In Higgs triplet models(HTM), the linear combination of neutral, charged and doubly charged fields of the scalar multiplets is linked to physical particles such as: heavy neutral boson, singly charged particles, doubly charged particles, etc. These physical particles contribute to many lepton number and lepton flavour violating processes, which can be probed at collider experiments. For detailed discussion, see Ref.\,\cite{Dziewit:2021pak}, and references therein.

	The model incorporates an additional symmetry, $Z_{10}$\,\cite{CarcamoHernandez:2020udg}, which is necessary to restrict some undesirable and next to leading order\,(NLO) terms in the Yukawa Lagrangian. A summary of all the field contents are shown in Table\,(\ref{table:1}).
\begin{largetable}
\caption{The transformation properties of the field contents under $ A_{4} \times Z_{10}\times SU(2)_{L}\times U(1)_Y$.}
\vspace{0.1cm}
\begin{tabular}{c|ccccccccccc}
\hline
Fields & $D_{l_{L}}$ &$l_{R}$ &$\nu_{l_R}$ & $\phi$ & $\Phi$ & $\eta$ & $\kappa$ & $\Delta$ & $\xi$ & $\zeta$ \\ 
\hline
$A_{4}$ & 3 & (1,\,1{$''$},\,1{$'$}) & (1,\,1{$''$},\,1{$'$}) & 3 & 3 & 1{$''$} & 1{$'$} & 3 & 1 & 1 \\
\hline
$Z_{10}$ & 0 & 0 & (0, 4, 6) & 0 & 0 & 2 & 8 & 0 & 6 & 4 \\
\hline
$SU(2)_{L}$ & 2 & 1 & 1 & 2 & 2 & 1 & 1 & 3 & 1 & 1\\
\hline
$U(1)_{Y}$ & -1 & -2 & 0 & 1 & -1 & 0 & 0 & -2 & 0 & 0\\
\hline
\end{tabular}
\label{table:1}
\end{largetable}

	The details of the basis and multiplication rules under $A_4$ symmetry can be found in the Refs\,\cite{Ma:2001dn, King:2006np, Altarelli:2010gt, King:2013eh}.
The Lagrangian in the leading order\,(LO) is formulated in the T basis\,\cite{Altarelli:2010gt} and it is presented in the following manner,
\begin{eqnarray}
- \mathcal{L}_Y &= y_e (\bar{D}_{l_L}\phi)\,e_{R} + y_{\mu} (\bar{D}_{l_L}\phi) \,\mu_{R}+\,y_{\tau} (\bar{D}_{l_L}\phi) \tau_{R}\nonumber\\&+ y_1\,(\bar{D}_{l_L}\Phi)\,\nu_{e_R} + \frac{y_2}{\Lambda}\,(\bar{D}_{l_L}\Phi) \,\nu_{{\mu}_{R}}\,\xi + \frac{y_3}{\Lambda}\,(\bar{D}_{l_L}\,\nonumber\\&\Phi)\,\nu_{{\tau}_{R}}\,\zeta +\,\frac{1}{2}\,M_1\,(\widebar{\nu^c_{e_R}}\,\nu_{e_R}) +\,\frac{1}{2}\,M_2\,[(\widebar{\nu^c_{\mu_{R}}}\nu_{\tau_{R}})\nonumber\\&+ (\widebar{\nu^c_{\tau_{R}}} \nu_{\mu_{R}})]+\,\frac{1}{2}\,y_{R_1}(\widebar{\nu^c_{\mu_{R}}}\,\nu_{\mu_{R}})\,\eta +\,\frac{1}{2}\,y_{R_2}(\widebar{\nu^c_{\tau_R}}\,\nonumber\\&\nu_{\tau_{R}})\,\kappa +\, y(\bar{D}_{l_L}\, D_{l_L}^c)\,{i \sigma_2 \Delta} \,+ h.c \quad\quad\quad\quad
\label{Yukawa Lagrangian}
\end{eqnarray}  
The vacuum alignments associated with the scalar fields $\phi$, $\Phi$ and $\Delta$ are taken as $\langle \phi \rangle = v(1,0,0)$, $\langle \Phi\rangle = u(0,1,1)$ and $\langle \Delta \rangle= w (0,1,-1)$, respectively. The vacuum alignments of the remaining scalar fields $\eta$, $\kappa$, $\xi$ and $\zeta$ are chosen as $\langle \eta \rangle  =\,v_{\eta}$, $\langle \kappa\rangle \,=\, v_{\kappa}$, $\langle \xi \rangle  =\,v_{\xi}$ and $ \langle \zeta\rangle \,=\, v_{\zeta}$. The scalar potential associated with the chosen alignments is depicted in Appendix(see eq.\ref{Pot}). The Lagrangian $\mathcal{L}_Y$ reveals the mass matrices as follows,
\begin{align}
& M_{l}=v \begin{bmatrix}
y_e & 0 & 0\\
0   & y_\mu & 0\\
0   &  0     & y_\tau
\end{bmatrix},\,
M_D = u\begin{bmatrix}
 0 & \widehat{y_2} & \widehat{y_3}\\
 y_1 & 0 & \widehat{y_3} \\
 y_1 & \widehat{y_2} & 0 \\
 \end{bmatrix}, &\nonumber\\
&\small M_R = \begin{bmatrix}
 M_1 & 0 & 0\\
 0 & M_3 & M_2 \\
 0 & M_2  & M_4\\
 \end{bmatrix},\,
\small M_{II} = \frac{y w}{3}\begin{bmatrix}
  0&1&-1\\
  1&2&0 \\
 -1&0&-2 \\
 \end{bmatrix},\nonumber\\&
\end{align}
where, $ y_{R_1} v_{\eta}=M_3$, $y_{R_2} v_{\kappa}=M_4$, $\widehat{y_2}= \frac{y_2 v_{\xi}}{\Lambda}$ and $\widehat{y_3}=\frac{y_3 v_{\zeta}}{\Lambda}$. The mass matrices $M_l$, $M_D$, $M_R$ and $M_{II}$ are attributed to the charged leptons, Dirac neutrinos, right handed Majorana neutrinos and the contribution arising from Type-II seesaw, respectively.
Considering the contribution arising from Type-I seesaw which is $M_I= -M_{D}\cdot M_{R}^{-1}\cdot M_{D}^{T}$, the light Majorana neutrino mass matrix ($M_{\nu}= M_{I}+ M_{II}$) takes the following form,
\begin{equation}
 M_{\nu} = 
 \begin{bmatrix}
 L + P & L & P\\
 L & G & J\\
 P & J  & H \\
 \end{bmatrix}, 
 \label{Mass matrix 5}
\end{equation} 
where, the \emph{texture parameters} viz., $L, P, G, J $ and $H$ are expressed in terms of the fifteen \emph{model parameters} viz., $y_1, y_2$,
 $y_3, y_{R_1}, y_{R_2}, y, u, v_{\eta}, v_{\kappa},  v_{\xi}, v_{\zeta}, w$, $\Lambda$, $M_1$ and $M_2$ as shown below\footnote{It is to be noted that texture parameters are not additional  parameters but functions of the model parameters. There is no linear transformation among the model to texture parameters. The latter is introduced to make the visualization simple.},
\begin{eqnarray}
\label{texture parameters1}
L &=&\frac{u^2 \hat{y_3} \left(\hat{y_3} y_{R_1} v_{\eta }-M_2 \hat{y_2}\right)}{M_2^2-y_{R_1} y_{R_2} v_{\eta } v_{\kappa }}+\frac{w y}{3},\\
P &=&\frac{u^2 \hat{y_2} \left(\hat{y_2} y_{R_2} v_{\kappa }-M_2 \hat{y_3}\right)}{M_2^2-y_{R_1} y_{R_2} v_{\eta } v_{\kappa }}-\frac{w y}{3},\\
G &=&u^2 \left(\frac{\hat{y_3}{}^2 y_{R_1} v_{\eta }}{M_2^2-y_{R_1} y_{R_2} v_{\eta } v_{\kappa }}-\frac{y_1^2}{M_1}\right)+\frac{2 w y}{3},\\
J&=&-u^2 \left(\frac{M_2 \hat{y_2} \hat{y_3}}{M_2^2-y_{R_1} y_{R_2} v_{\eta } v_{\kappa }}+\frac{y_1^2}{M_1}\right),\\
H &=&u^2 \left(\frac{\hat{y_2}{}^2 y_{R_2} v_{\kappa }}{M_2^2-y_{R_1} y_{R_2} v_{\eta } v_{\kappa }}-\frac{y_1^2}{M_1}\right)-\frac{2 w y}{3}.
\label{texture parameters2}
\end{eqnarray}
The texture appearing in eq.\,(\ref{Mass matrix 5}) bears a distinct feature in terms of a sum rule
\begin{equation}
m_{ee}= m_{e\mu}+m_{e\tau}.
\end{equation}

If we want the \emph{realistic} mixing matrix proposed in eq.\,(\ref{PMNS1}) to diagonalize $M_{\nu}$ in eq.\,(\ref{Mass matrix 5}),
\begin{eqnarray}
\label{diag}
\tilde{U}^T \cdot M_{\nu} \cdot \tilde{U} = diag\,(m_1\,e^{-2 i \alpha}, m_2\,e^{-2 i \beta}, m_3 ),
\end{eqnarray}
then, we see there will appear seven constraints which will further restrict the number of free texture parameters to three which are $Re[P]$, $Im[P]$ and $Re[G]$.
\section{Numerical Analysis and Important Observations}
\label{section4}
\begin{figure*}
  \centering
    \subfigure[]{\includegraphics[width=0.2545
  \textwidth]{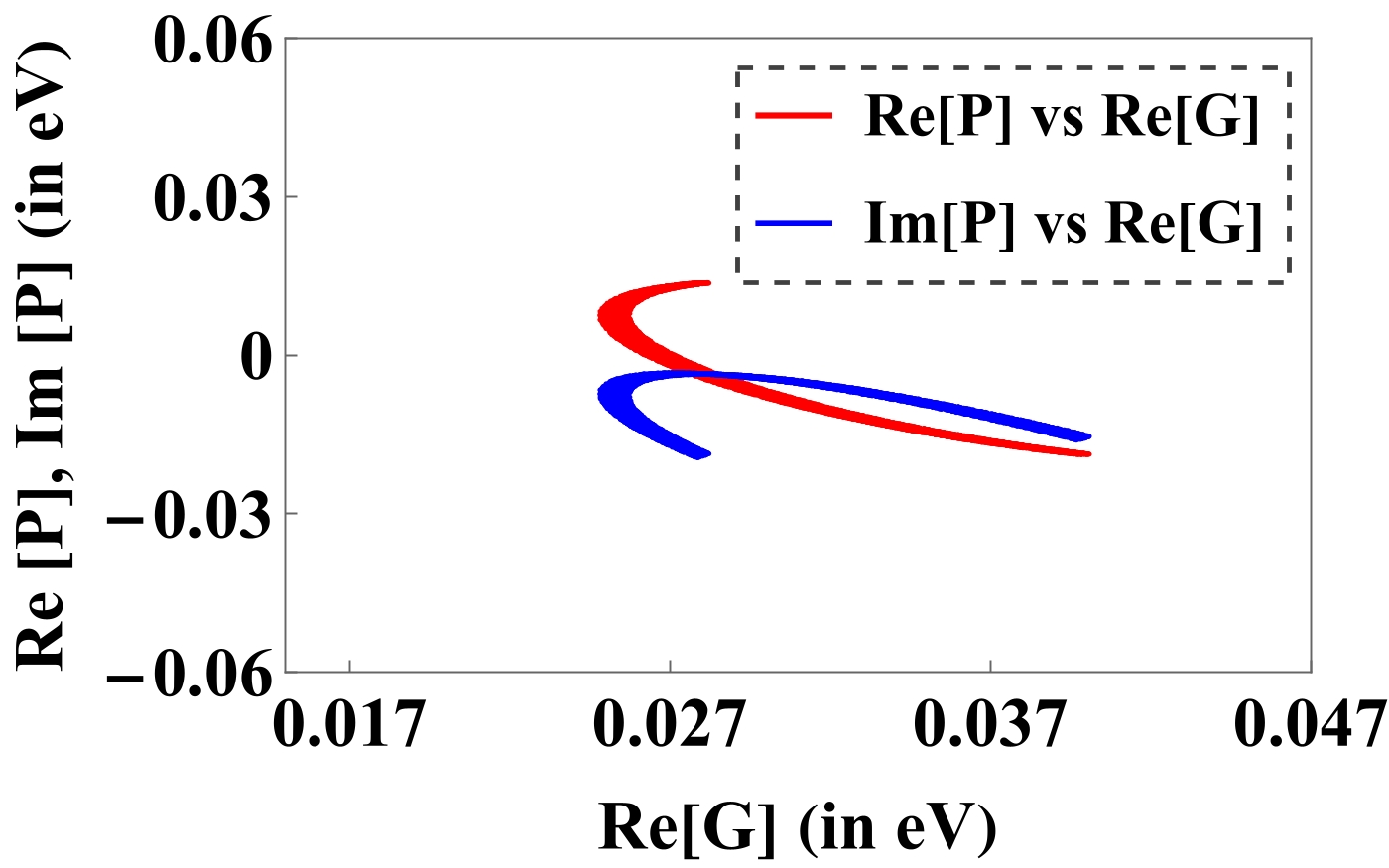}\label{fig:1(a)}} 
    \subfigure[]{\includegraphics[width=0.2435\textwidth]{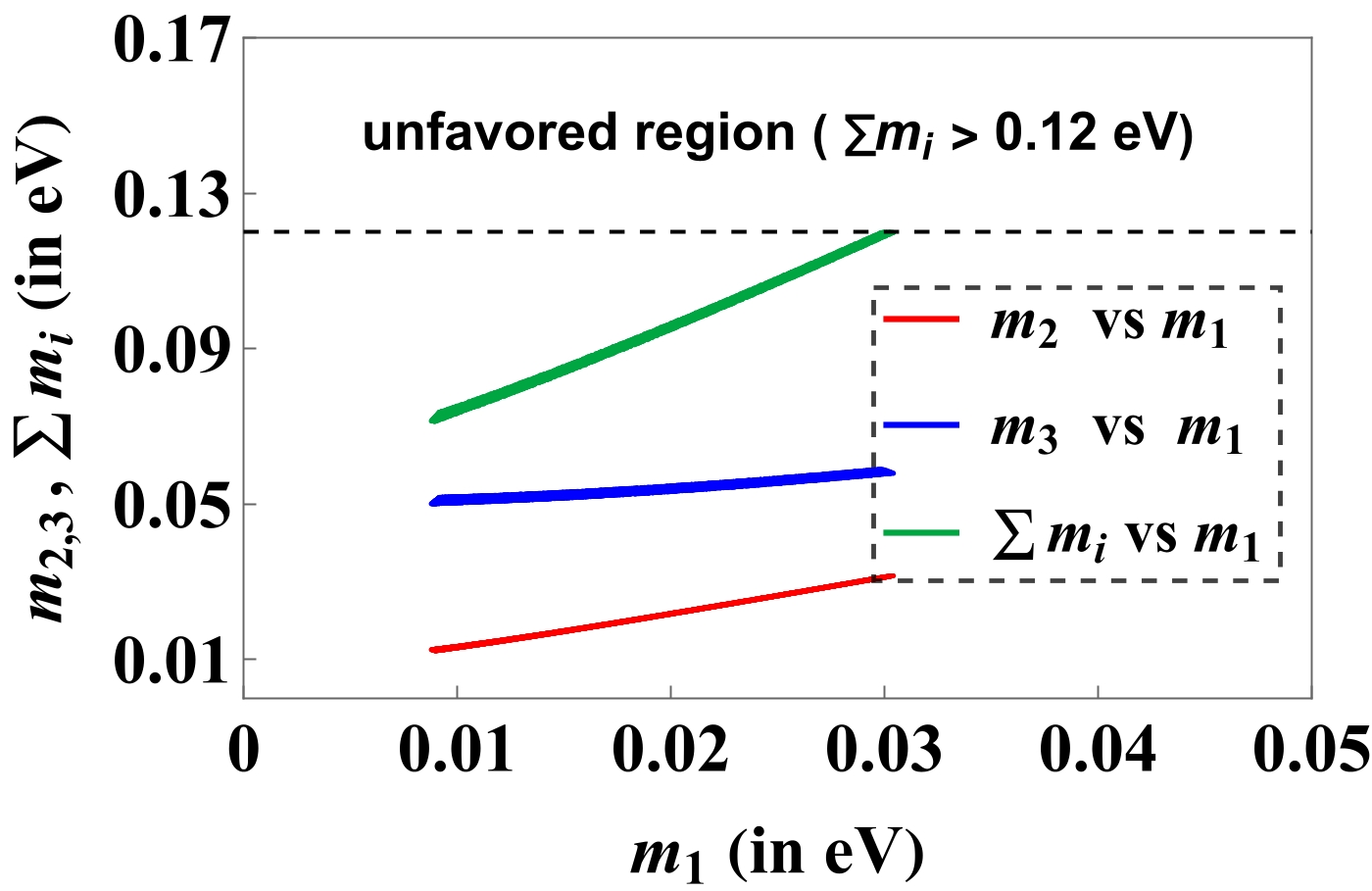}\label{fig:1(b)}} 
    \subfigure[]{\includegraphics[width=0.24\textwidth]{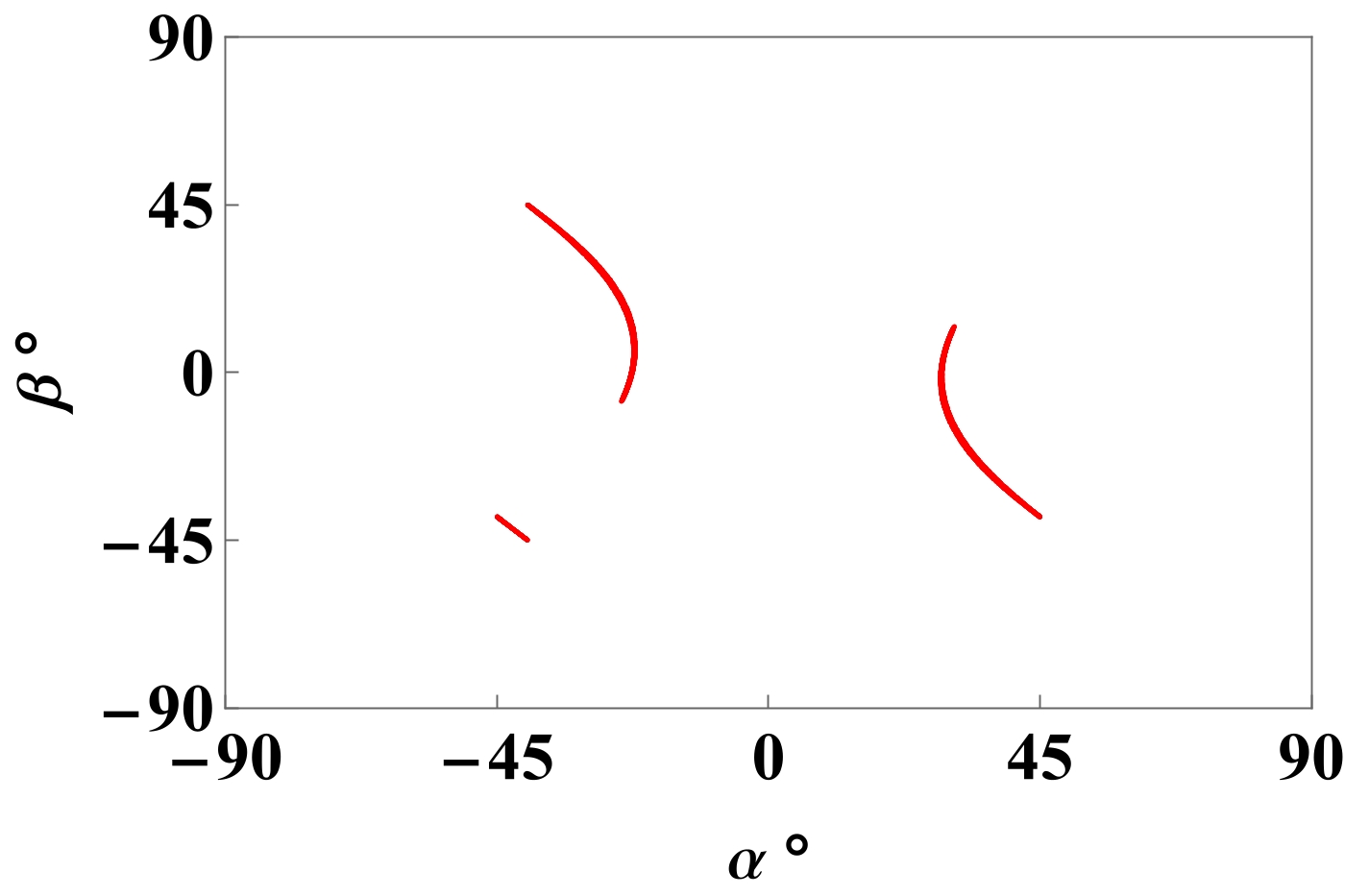}\label{fig:1(c)}}
     \subfigure[]{\includegraphics[width=0.242
  \textwidth]{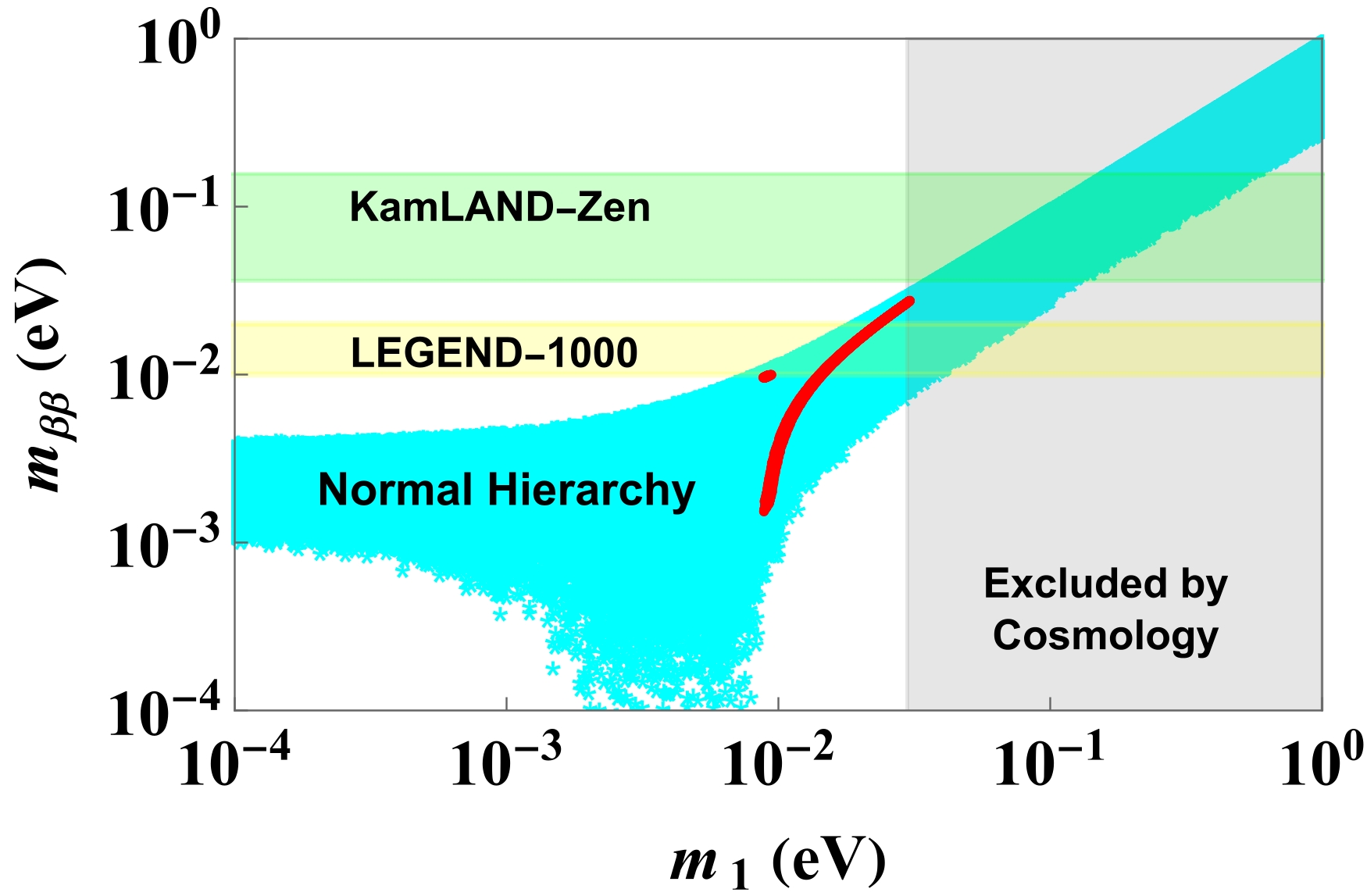}\label{fig:1(d)}}
\caption{The correlation plots for the free parameters, mass eigenvalues, Majorana phases and $m_{\beta\beta}$ are shown: (a)Represents the plots between the free parameters: i.e $Re[P]$ vs $Re[G]$ and $Im[P]$ vs $Re[G]$.(b) Highlights the plots of $m_2$ vs $m_1$, $m_3$ vs $m_1$ and $m_1+m_2+m_3$ vs $m_1$.(c) Shows the plot between Majorana phases $\alpha$ and $\beta$. (d) Represents the plot between the effective Majorana neutrino mass $m_{\beta\beta}$ and the lightest neutrino mass eigenvalue $m_1$. Here, the `cyan' band shows the region allowed by current oscillation data. `Green' and `yellow' bands represent KamLAND-Zen and LEGEND-1000 sensitivities. The `gray' band is the region excluded by cosmology for the normal hierarchy, and the `red' plot is our model's prediction.}
\label{fig:1}
\end{figure*}

\begin{figure*}
  \centering
    \subfigure[]{\includegraphics[width=0.243
  \textwidth]{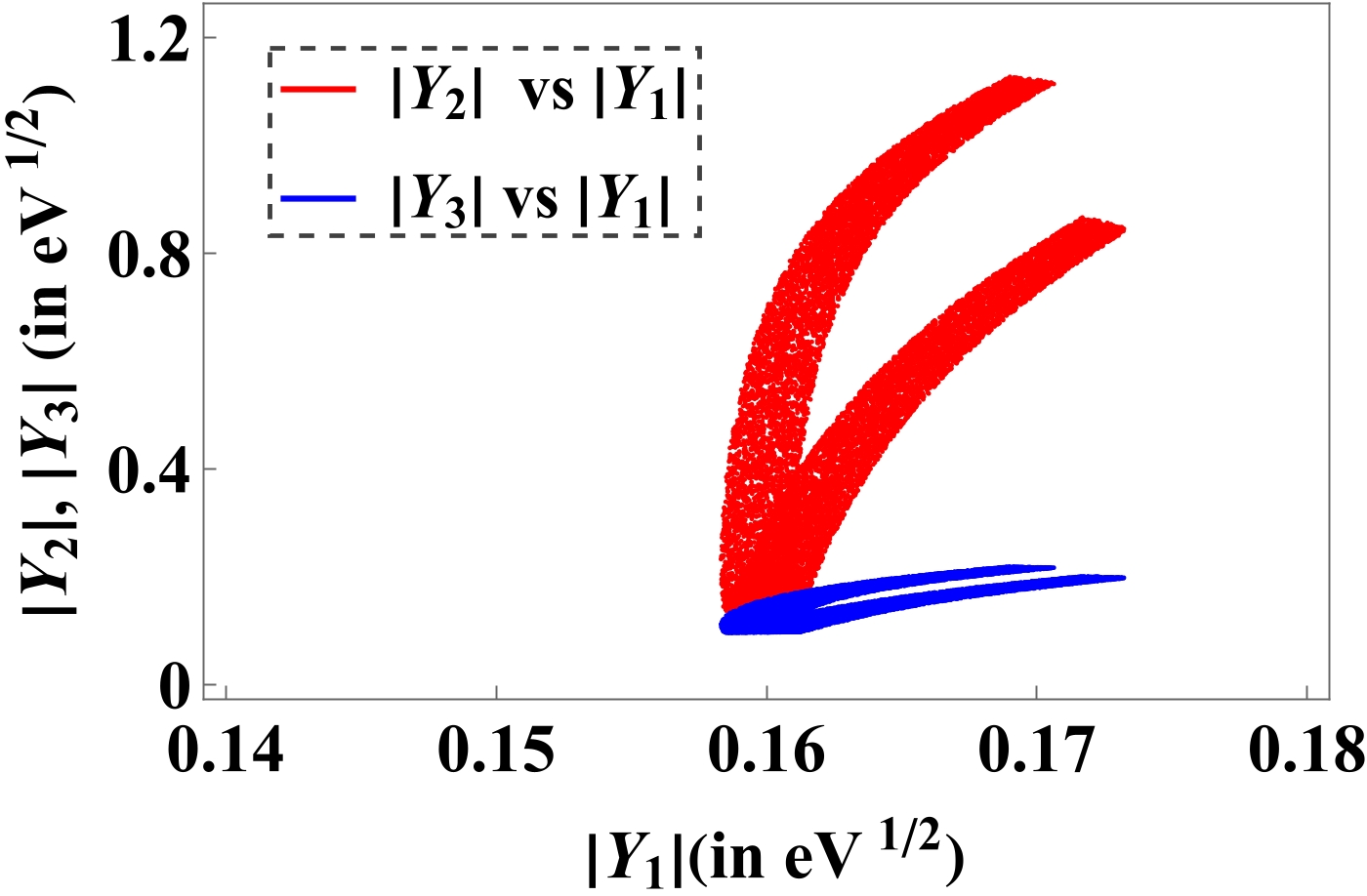}\label{fig:2(a)}} 
    \subfigure[]{\includegraphics[width=0.243\textwidth]{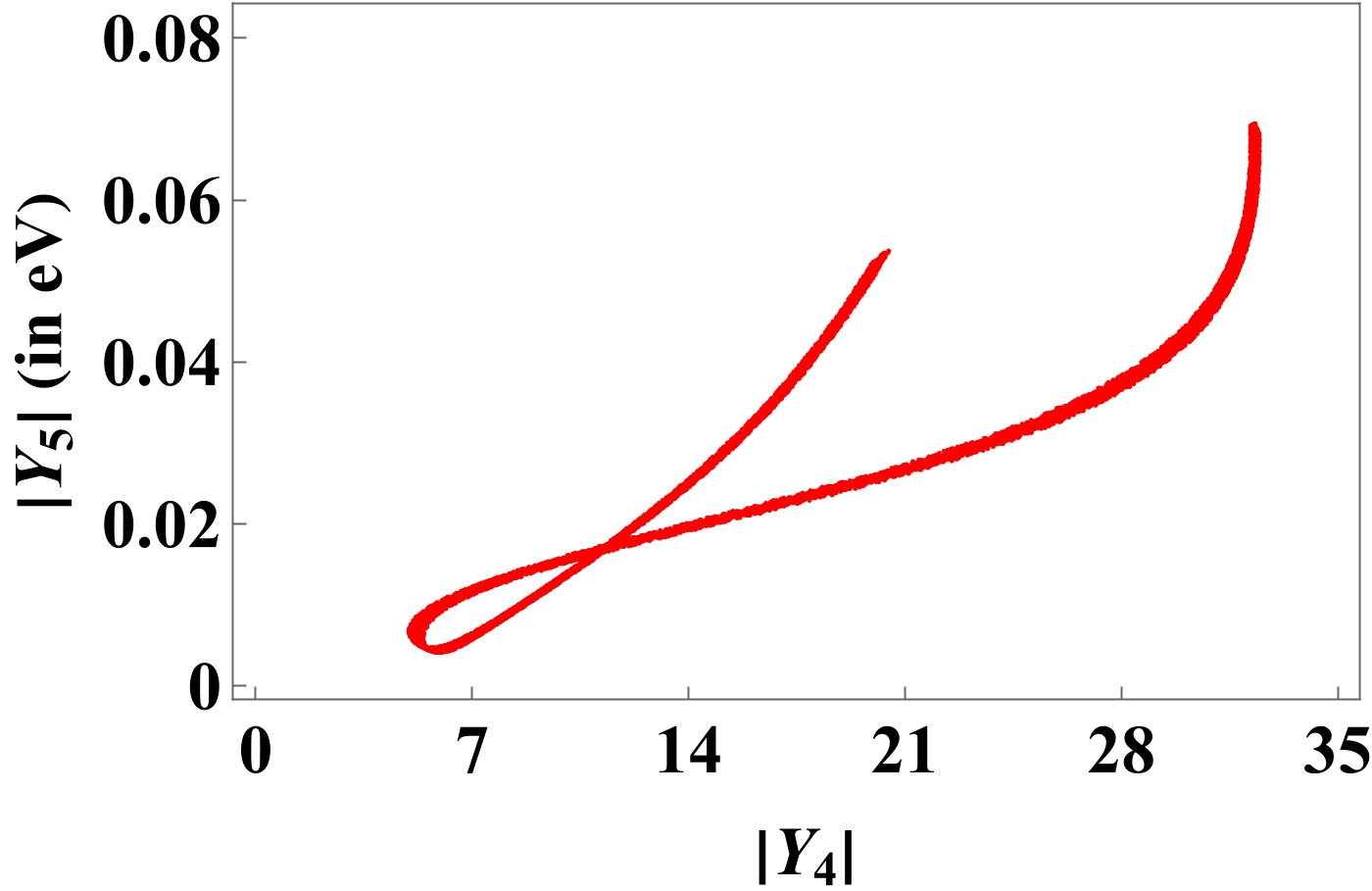}\label{fig:2(b)}} 
    \subfigure[]{\includegraphics[width=0.2428\textwidth]{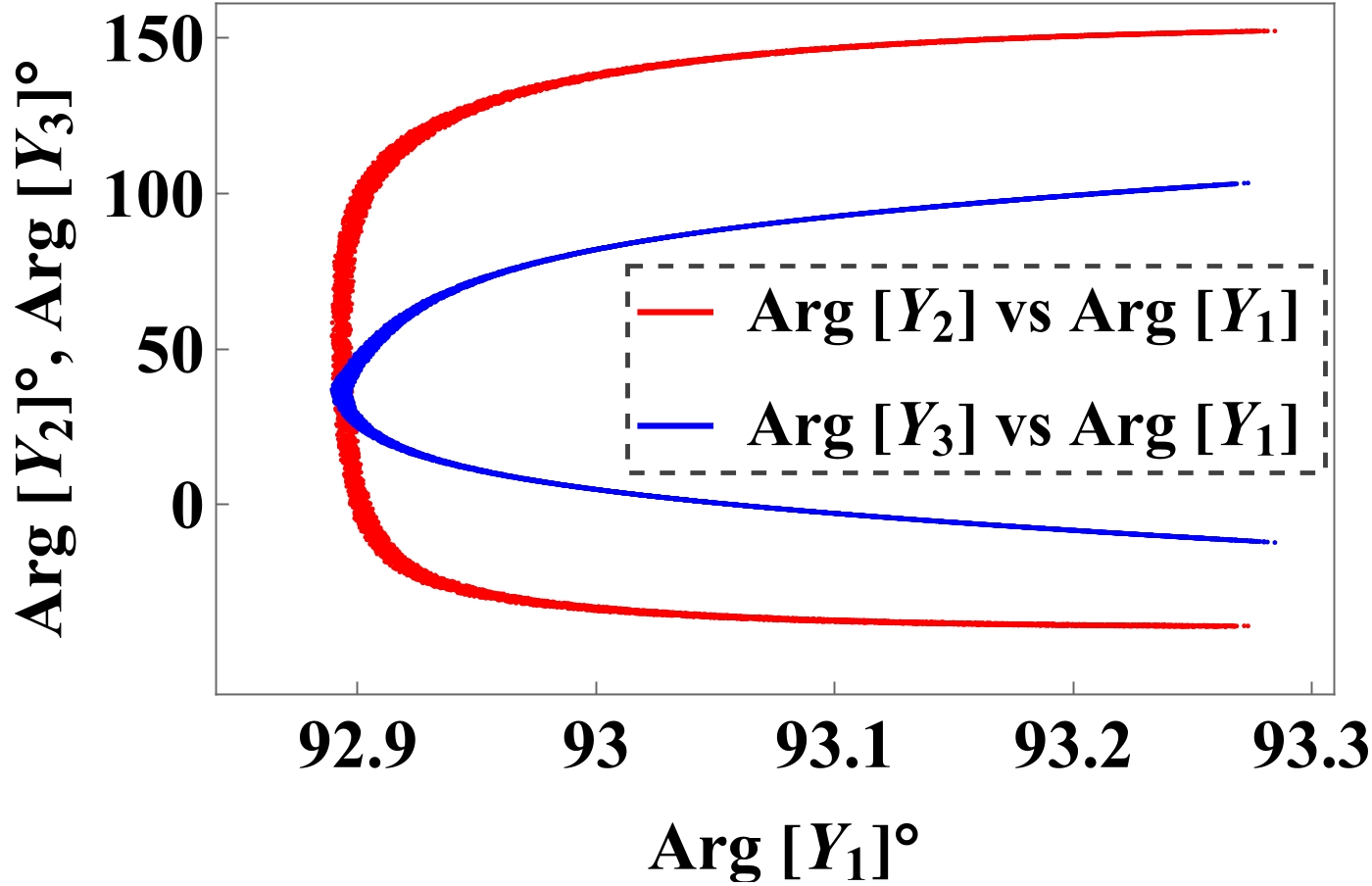}\label{fig:2(c)}}
    \subfigure[]
{\includegraphics[width=0.251\textwidth]{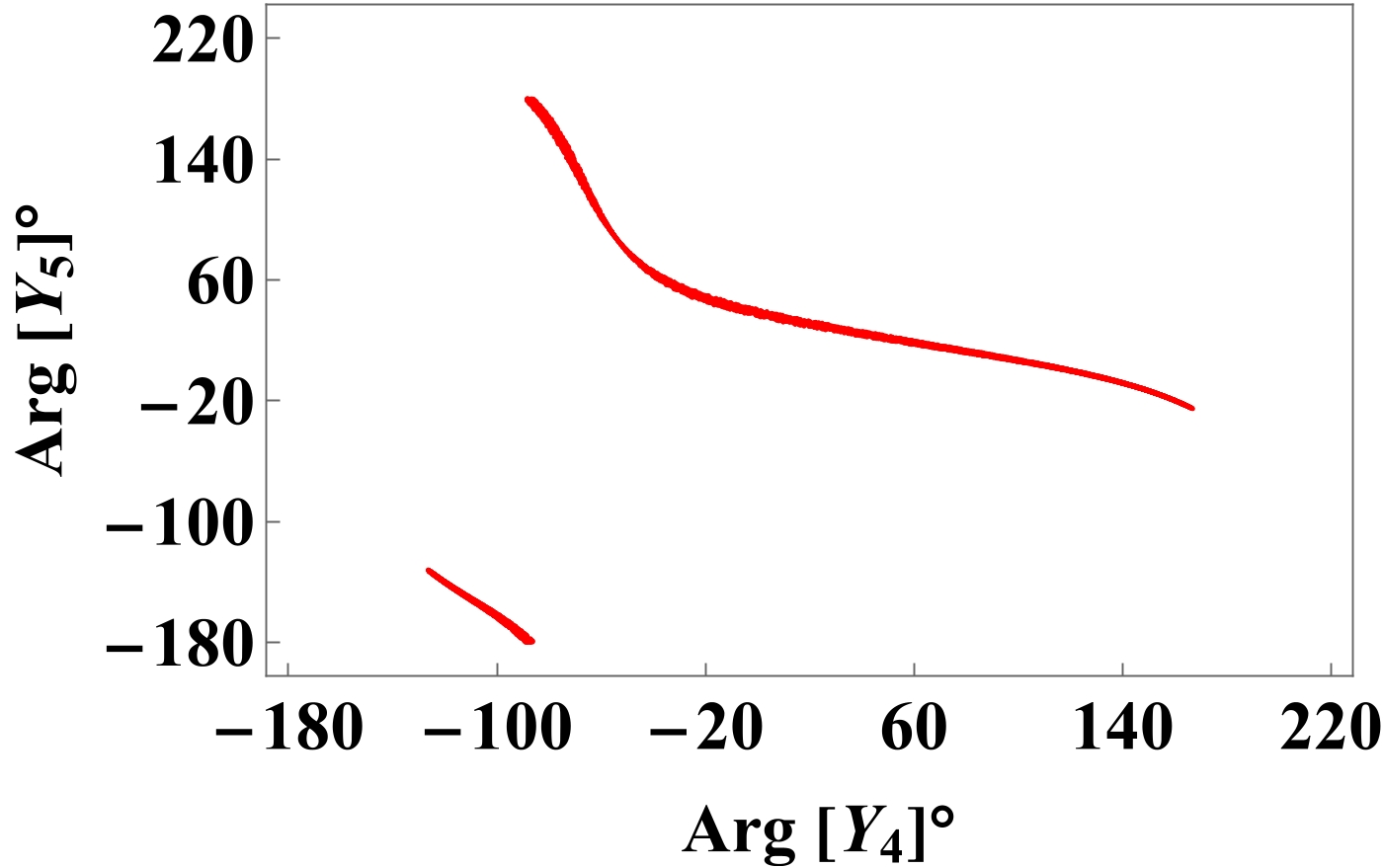}\label{fig:2(d)}}
\caption{The correlation plots for the model parameters are manifested: (a)Represents the plots of $|Y_2|$ vs $|Y_1|$  and $|Y_3|$ vs $|Y_1|$. (b) Depicts the plot between $|Y_4|$ and $|Y_5|$. (c) Shows the plots of Arg[$Y_2$] vs Arg[$Y_1$] and Arg[$Y_3$] vs Arg[$Y_1$]. (d) Highlights the plot between Arg[$Y_4$] and Arg[$Y_5$].}
\label{fig:2}
\end{figure*}

We see that the three mass eigenvalues and two Majorana phases are expressible in terms of $Re[P]$, $Im[P]$ and $Re[G]$. We generate sufficient number of data points $\lbrace Re[P],Im[P],Re[G]\rbrace$ such that the observational parameters $\Delta m_{21}^2$ and $\Delta m_{31}^2$ lie within the $3\sigma$ bounds\,\cite{Gonzalez-Garcia:2021dve} and the sum of three neutrino masses, $m_1+m_2+m_3 <0.12\,\text{eV}$\,\cite{Planck:2018vyg}. However, the similar analysis when conducted for inverted ordering does not give rise to any fruitful result.
From the datasets we see that,
\begin{gather*}
\begin{aligned}
(Re[P])_{\text{min}} & = -0.0187\,\text{eV}, & (Re[P])_{\text{max}} & = 0.0138\,\text{eV}; \\
(Im[P])_{\text{min}} & = -0.0194\,\text{eV}, & (Im[P])_{\text{max}} & = -0.0033\,\text{eV}; \\
(Re[G])_{\text{min}} & = 0.0248\,\text{eV},  &  (Re[G])_{\text{max}} & = 0.0400\,\text{eV}.
\end{aligned}
\end{gather*}
The datasets concerning the free parameters are graphically visualised in Fig.\,\ref{fig:1(a)}. 
The three mass eigenvalues i.e, $m_1$, $m_2$, $m_3$, the sum of three mass eigenvalues i.e, $\sum{m_i}$\,($i= 1,2,3$), and the two Majorana CP phases, $\alpha$ and $\beta$, are seen to exhibit their lowest and highest values as manifested in the Table\,\ref{table:2}. In addition, we try to see what inference we may draw regarding the effective Majorana neutrino mass, $m_{\beta\beta}= |U^2_{ei} m_i|$\,($i= 1,2,3$) which is an observational parameter in neutrinoless double beta\,($0\nu\beta\beta$)-decay\,\cite{Barabash:2023dwc, Agostini:2022zub}. So far, we are aware of the upper bounds concerning this parameter which are $<(75-350)\,meV$\,\cite{CUORE:2019yfd}, $<(104-228)\,meV$\,\cite{GERDA:2019ivs} and $<(61-165)\,meV$\,\cite{KamLAND-Zen:2016pfg}. In Table\ref{table:2}, the minimum and maximum values of $m_{\beta\beta}$ are predicted from our model, and these are in agreement with experiments.
We graphically represent these predictions in figs.\,\ref{fig:1(b)} - \ref{fig:1(d)}.
\begin{table}[h]
\caption{ Represents the lowest and highest values of the mass eigenvalues $m_1$, $m_2$, $m_3$, the sum of mass eigenvalues $\sum{m_i}$, the Majorana phases $\alpha$, $\beta$ and the effective Majorana neutrino mass $m_{\beta\beta}$.}
\centering
\begin{tabular}{lcc}
\hline
\hline
Prediction & Lowest Value & Highest Value\\
\hline
\hline
$m_1 /\text{eV}$ & 0.0087 & 0.0304\\
\hline
$m_2 /\text{eV}$ & 0.0122 & 0.0315 \\
\hline
$m_3 /\text{eV}$ & 0.0500 & 0.0590\\
\hline
$\sum{m_i} /\text{eV}$ &0.0712 &0.1199\\
\hline
$\alpha$ & $-45^{\circ}$ & $45^{\circ}$\\
\hline
$\beta$ & $-45^{\circ}$ & $45^{\circ}$ \\
\hline
$m_{\beta\beta}/\text{meV}$ &1.5570& 28.0442\\
\hline
\end{tabular} 
\label{table:2}
\end{table}
From equations\,(\ref{texture parameters1}-\ref{texture parameters2}), we express the texture parameters in terms of the model parameter combinations $Y_1, Y_2, Y_3, Y_4$ and $Y_5$ as shown in the Appendix\,(see eq. \ref{mpc}). Here, $\frac{u y_1}{\sqrt{M_1}}=Y_1$, $\frac{u y_2 v_{\xi } \sqrt{y_{R_2} v_{\kappa }}}{\Lambda  M_2}=Y_2$, $\frac{u y_3 v_{\zeta }}{\Lambda  \sqrt{y_{R_2} v_{\kappa }}}=Y_3$, $\frac{y_{R_1} y_{R_2} v_{\eta } v_{\kappa }}{M_2^2}=Y_4$ and $w\,y=Y_5$. It is noteworthy that the model demonstrates accurate predictions for these specific combinations of model parameters. The predicted maximum and minimum values of the latter are presented in Table \ref{table:3}.
\begin{table}[h]
\caption{ Represents the maximum and minimum values of the model parameters.}
\centering
\begin{tabular}{ccc}
\hline
\hline
Parameter & Minimum Value & Maximum Value\\
\hline
\hline
$|Y_1|/\,\text{eV}^{\frac{1}{2}}$ & 0.1583 & 0.1732\\
\hline
$|Y_2|/\,\text{eV}^{\frac{1}{2}}$ & 0.1364 & 1.1276 \\
\hline
$|Y_3|/\,\text{eV}^{\frac{1}{2}}$ & 0.0971 & 0.2192\\
\hline
$|Y_4|$ & 4.9486 & 32.4693\\
\hline
$|Y_5|/ \text{eV} $ & 0.0041 & 0.0694\\ 
\hline
$Arg[Y_1]/^{\circ}$ &92.8892 &93.2845\\
\hline
$Arg[Y_2]/^{\circ}$ &-39.2312 & 152.2550\\
\hline
$Arg[Y_3]/^{\circ}$ &-12.2087 & 103.3790\\
\hline
$Arg[Y_4]/^{\circ}$ &-126.4170 & 166.8890\\
\hline
$Arg[Y_5]/^{\circ}$ &-179.9890 & 179.9670\\
\hline
\end{tabular} 
\label{table:3}
\end{table}
The respective plots for the model parameters are presented in figs.\ref{fig:2(a)}- \ref{fig:2(d)}. 
\section{Summary and Discussion}
\label{section5}
In summary, we have hoisted a \emph{realistic} neutrino mixing scheme which is consistent with the experiments highlighting, $\theta_{12}\approx 35.26^\circ$, $\theta_{13}\approx 8.29^\circ$, $\theta_{23}\approx 48.18^\circ$ and $\delta = 270^\circ$. The choice of this mixing scheme is special in the sense that it can be realised in a Type-I+Type-II seesaw model based on $A_4 \times Z_{10}$. The additional symmetry, $Z_{10}$, restricts some undesirable terms in the Yukawa Lagrangian. The model is not hierarchy blind and obeys the normal ordering of neutrino masses. It highlights that the minimum/maximum value which the two Majorana phases can take is $-45^\circ/45^\circ$. In the light of the present model, we have predicted the maximum value of $m_{\beta\beta}$ as 28.0442 meV which lies below the experimental upper bound.   
If we look independently into the texture appearing in equation\,(\ref{Mass matrix 5}), then we see that it holds five independent complex \emph{texture} parameters which are expressed in terms of fifteen \emph{model} parameters. If we want the mass matrix to be diagonalised under the proposed mixing scheme, we see that only three real texture parameters exist which in turn, helps us in dealing with the complexities associated with the model parameters effectively.

	The present model is equipped with the numerical information related to the model parameters and hence it can be employed to study the lepton flavour violation and baryon asymmetry of the Universe. This happens to be our future prospect. 
\acknowledgments
The Research work of Manash Dey is supported  by the Council of Scientific and Industrial Research\,(CSIR), Government of India through a NET Junior Research Fellowship vide grant No. 09/0059(15346)/2022-EMR-I.
\subsection{Data availability statement:}
The data that support the findings of this study are available upon reasonable request from the authors.

\
\bibliographystyle{eplbib} 
\bibliography{reference1.bib}

\begin{thebibliography}{10}
\expandafter\ifx\csname url\endcsname\relax\def\url#1{\texttt{#1}}\fi

\bibitem{Cowan:1956rrn}
\Name{Cowan C.~L., Reines F., Harrison F.~B., Kruse H.~W. \and McGuire A.~D.}
  \REVIEW{Science}{124}{1956}{103}.

\bibitem{SNO:2002tuh}
\Name{Ahmad Q.~R. \etal} \REVIEW{Phys. Rev. Lett.}{89}{2002}{011301}.

\bibitem{KamLAND:2002uet}
\Name{Eguchi K. \etal} \REVIEW{Phys. Rev. Lett.}{90}{2003}{021802}.

\bibitem{Super-Kamiokande:1998uiq}
\Name{Fukuda Y. \etal} \REVIEW{Phys. Rev. Lett.}{82}{1999}{2644}.

\bibitem{Pontecorvo:1957cp}
\Name{Pontecorvo B.} \REVIEW{Sov. Phys. JETP}{6}{1957}{429}.

\bibitem{Ramond:1993kv}
\Name{Ramond P., Roberts R.~G. \and Ross G.~G.} \REVIEW{Nucl. Phys.
  B}{406}{1993}{19}.

\bibitem{Ibarra:2003Jb}
\Name{Ibarra A. \and Ross G.} \REVIEW{PoSA}{HEP2003}{2003}{030}.

\bibitem{Harrison:2002er}
\Name{Harrison P.~F., Perkins D.~H. \and Scott W.~G.} \REVIEW{Phys. Lett.
  B}{530}{2002}{167}.

\bibitem{Dey:2022qpu}
\Name{Dey M., Chakraborty P. \and Roy S.} \REVIEW{Phys. Lett.
  B}{839}{2023}{137767}.

\bibitem{Xing:2022uax}
\Name{Xing Z.-z.} \REVIEW{Rept. Prog. Phys.}{86}{2023}{076201}.

\bibitem{Xing:2015fdg}
\Name{Xing Z.-z. \and Zhao Z.-h.} \REVIEW{Rept. Prog. Phys.}{79}{2016}{076201}.

\bibitem{Xing:2020ijf}
\Name{Xing Z.-z.} \REVIEW{Phys. Rept.}{854}{2020}{1}.

\bibitem{Ohlsson:2013xva}
\Name{Ohlsson T. \and Zhou S.} \REVIEW{Nature Commun.}{5}{2014}{5153}.

\bibitem{King:2003jb}
\Name{King S.~F.} \REVIEW{Rept. Prog. Phys.}{67}{2004}{107}.

\bibitem{Cai:2017mow}
\Name{Cai Y., Han T., Li T. \and Ruiz R.} \REVIEW{Front. in
  Phys.}{6}{2018}{40}.

\bibitem{Mohapatra:2004zh}
\Name{Mohapatra R.~N.} \Book{{Seesaw mechanism and its implications}} in proc.
  of \Book{{SEESAW25: International Conference on the Seesaw Mechanism and the
  Neutrino Mass}} 2004 pp. 29--44.

\bibitem{Cheng:1980qt}
\Name{Cheng T.~P. \and Li L.-F.} \REVIEW{Phys. Rev. D}{22}{1980}{2860}.

\bibitem{Wong:2022qyg}
\Name{Wong C.-F. \and Chen Y.} \REVIEW{Phys. Lett. B}{833}{2022}{137354}.

\bibitem{Akhmedov:2006de}
\Name{Akhmedov E.~K. \and Frigerio M.} \REVIEW{JHEP}{01}{2007}{043}.

\bibitem{Maki:1962mu}
\Name{Maki Z., Nakagawa M. \and Sakata S.} \REVIEW{Prog. Theor.
  Phys.}{28}{1962}{870}.

\bibitem{Barger:1998ta}
\Name{Barger V.~D., Pakvasa S., Weiler T.~J. \and Whisnant K.} \REVIEW{Phys.
  Lett. B}{437}{1998}{107}.

\bibitem{Albright:2010ap}
\Name{Albright C.~H., Dueck A. \and Rodejohann W.} \REVIEW{Eur. Phys. J.
  C}{70}{2010}{1099}.

\bibitem{Ding:2011cm}
\Name{Ding G.-J., Everett L.~L. \and Stuart A.~J.} \REVIEW{Nucl. Phys.
  B}{857}{2012}{219}.

\bibitem{Gonzalez-Garcia:2021dve}
\Name{Gonzalez-Garcia M.~C., Maltoni M. \and Schwetz T.}
  \REVIEW{Universe}{7}{2021}{459}.

\bibitem{Ma:2001dn}
\Name{Ma E. \and Rajasekaran G.} \REVIEW{Phys. Rev. D}{64}{2001}{113012}.

\bibitem{Dziewit:2021pak}
\Name{Dziewit B., Kordiaczy\'nska M. \and Srivastava T.}
  \REVIEW{Symmetry}{13}{2021}{1240}.

\bibitem{CarcamoHernandez:2020udg}
\Name{C\'arcamo~Hern\'andez A.~E. \and de~Medeiros~Varzielas I.} \REVIEW{Phys.
  Lett. B}{806}{2020}{135491}.

\bibitem{King:2006np}
\Name{King S.~F. \and Malinsky M.} \REVIEW{Phys. Lett. B}{645}{2007}{351}.

\bibitem{Altarelli:2010gt}
\Name{Altarelli G. \and Feruglio F.} \REVIEW{Rev. Mod. Phys.}{82}{2010}{2701}.

\bibitem{King:2013eh}
\Name{King S.~F. \and Luhn C.} \REVIEW{Rept. Prog. Phys.}{76}{2013}{056201}.

\bibitem{Planck:2018vyg}
\Name{Aghanim N. \etal} \REVIEW{Astron. Astrophys.}{641}{2020}{A6} [Erratum:
  Astron.Astrophys. 652, C4 (2021)].

\bibitem{Barabash:2023dwc}
\Name{Barabash A.} \REVIEW{Universe}{9}{2023}{290}.

\bibitem{Agostini:2022zub}
\Name{Agostini M., Benato G., Detwiler J.~A., Men\'endez J. \and Vissani F.}
  \REVIEW{Rev. Mod. Phys.}{95}{2023}{025002}.

\bibitem{CUORE:2019yfd}
\Name{Adams D.~Q. \etal} \REVIEW{Phys. Rev. Lett.}{124}{2020}{122501}.

\bibitem{GERDA:2019ivs}
\Name{Agostini M. \etal} \REVIEW{Science}{365}{2019}{1445}.

\bibitem{KamLAND-Zen:2016pfg}
\Name{Gando A. \etal} \REVIEW{Phys. Rev. Lett.}{117}{2016}{082503} [Addendum:
  Phys.Rev.Lett. 117, 109903 (2016)].

\end{thebibliography}
\section{ Appendix}
\subsection{Parametrisation of the PMNS matrix}
The parametrisation of the PMNS matrix\,($U$) adopted by the Particle Data Group is shown below,
\begin{equation}
\label{pmns}
U= P_{\phi}\cdot \tilde{U} \cdot P_M ,
\end{equation}
where, $P_{\phi}=diag (e^{i\phi_1},e^{i\phi_2},e^{i\phi_3})$ and $P_M=diag(e^{i\alpha},e^{i\beta},1)$. In principle, the three arbitrary phases\,($\phi_{i=1,2,3}$) can be eliminated from $U$ by redefining the charged lepton fields in terms of these phases. Hence, the PMNS matrix takes the following form,
\begin{equation}
\label{pmns1}
U= \tilde{U} \cdot P_M .
\end{equation} 
The matrix $\tilde{U}$ is represented as shown in the following,
\begin{eqnarray}
\tilde{U} &=& \begin{bmatrix}
1 & 0 & 0\\
0 & c_{23} & s_{23}\\
0 & - s_{23} & c_{23}
\end{bmatrix}\times \begin{bmatrix}
c_{13} & 0 & s_{13}\,e^{-i\delta}\\
0 & 1 & 0\\
-s_{13} e^{i\delta} & 0 & c_{13}
\end{bmatrix}\nonumber\\
&& \quad\quad\quad\times\begin{bmatrix}
c_{12} & s_{12} & 0\\
-s_{12} & c_{12} & 0\\
0 & 0 & 1
\end{bmatrix},
\end{eqnarray}
where, $s_{ij}=\sin\theta_{ij}$ and $c_{ij}=\cos\theta_{ij}$. 
In the basis where the charged lepton mass matrix is diagonal, then in general $U$ is the matrix that diagonalizes $M_{\nu}$, i.e,
\begin{equation}
U^T \cdot M_{\nu}\cdot U= diag(m_1, m_2, m_3),
\label{11}
\end{equation}
Using equation\,(\ref{pmns1}) in equation\,(\ref{11}) one obtains,
\begin{equation}
(\tilde{U} \cdot P_M)^T \cdot M_{\nu}\cdot\,(\tilde{U}\cdot P_M)= diag(m_1, m_2, m_3).
\label{12}
\end{equation}
We rearrange the above equation in the following way,
\begin{equation}
\tilde{U}^T \cdot M_{\nu}\cdot\,\tilde{U}= diag(\tilde{m_1},\tilde{m_2},m_3),
\label{13}
\end{equation}
where $\tilde{m_1} = m_1 e^{-2 i \alpha}$ and $\tilde{m_2} = m_2 e^{-2 i \beta}$. Now, $\tilde{U}$ is the observational matrix/PMNS matrix which diagonalises the neutrino mass matrix $M_{\nu}$.
\subsection{Model parameters in terms of texture parameters}
 Using eqs.\,(\ref{texture parameters1}-\ref{texture parameters2}), the model parameters are expressed in terms of the texture parameters as shown below:
\begin{align}
\label{mpc}
&Y_1=\frac{1}{2}(-G-H-2 J+L+P)^{\frac{1}{2}},\\
&Y_2=((G^2+G (-2 H-5 L+3 P)+H^2+3 H L-5 H P\nonumber\\&\quad\quad\,+2 J L+2 J P+4 L^2-8 L P+4 P^2)(3 G-H-2 J\nonumber\\&\quad\quad\,-3 L+5 P))^{\frac{1}{2}}/(G+H-2 J-L-P),\\
&Y_3=(G^2+G (-2 H-5 L+3 P)+H^2+3 H L-5 H P\nonumber\\&\quad\quad\,+2 J L+2 J P+4 L^2-8 L P+4 P^2)^{\frac{1}{2}}/(3 G-H-\nonumber\\&\quad\quad\,2 J-3 L+5 P)^{\frac{1}{2}},\\
&Y_4=-(3 G^2-2 G (5 H-2 J+9 L-7 P)+3 H^2+2 H \nonumber\\&\quad\quad\,(2 J+7 L-9 P)-4 J^2+4 J L+4 J P+15 L^2-\nonumber\\&\quad\quad\,34 L P+15 P^2)/(-G-H+2 J+L+P)^2,\\
&Y_5=(3 (G - H - L + P))/2 .
\end{align}
In the above equations, for the sake of simplicity, we redefine the model parameter combinations as:  $u y_1/\sqrt{M_1}=Y_1$, $u\,y_2 v_{\xi } \sqrt{y_{R_2} v_{\kappa }}/(\Lambda  M_2)=Y_2$, $u\,y_3 v_{\zeta }/(\Lambda  \sqrt{y_{R_2} v_{\kappa }})=Y_3$, $y_{R_1} y_{R_2} v_{\eta } v_{\kappa }/M_2^2=Y_4$ and $w\,y=Y_5$. The texture parameters namely $L, P, G, J $ and $H$ are dependent on the three independent real input parameters $Re[P], Im[P]$ and $Re[G]$. Therefore, the latter spans the model parameters as well. 
\subsection{Scalar Potential}
The scalar potential invariant under $ A_{4} \times Z_{10}\times SU(2)_{L}\times U(1)_Y$ is presented as shown:
\begin{eqnarray}
\label{Pot}
V&= V(\phi)+ V(\Phi)+V(\Delta)+V(\eta)+ V(\kappa)+V(\xi)\nonumber\\&\quad+V(\zeta)+V(interaction).
\end{eqnarray}
Writing the terms explicitly, we have,
\begin{align}
V(\phi)&= -\mu^2_{\phi}(\phi^{\dagger}\phi)+ \lambda^{\phi}_1(\phi^{\dagger}\phi)(\phi^{\dagger}\phi)+\lambda^{\phi}_2(\phi^{\dagger}\phi)_{1'}\nonumber\\&\quad\,(\phi^{\dagger}\phi)_{1''}+\lambda^{\phi}_3(\phi^{\dagger}\phi)_{3_s}(\phi^{\dagger}\phi)_{3_s}+\lambda^{\phi}_4(\phi^{\dagger}\phi)_{3_s}\nonumber\\&\quad\,(\phi^{\dagger}\phi)_{3_a}+\lambda^{\phi}_5(\phi^{\dagger}\phi)_{3_a}(\phi^{\dagger}\phi)_{3_a},\nonumber\\
V(\Phi)&= V(\phi \rightarrow \Phi),\nonumber\\
V(\Delta)&=-\mu^2_{\Delta}Tr(\Delta^{\dagger}\Delta)+ \lambda^{\Delta}_1 Tr(\Delta^{\dagger}\Delta)Tr(\Delta^{\dagger}\Delta)\nonumber\\&\quad\,+\lambda^{\Delta}_2 Tr(\Delta^{\dagger}\Delta)_{1'} Tr(\Delta^{\dagger}\Delta)_{1''}+\lambda^{\Delta}_3 Tr(\Delta^{\dagger}\nonumber\\&\quad\,\,\Delta)_{3_s} Tr(\Delta^{\dagger}\Delta)_{3_s}+\lambda^{\Delta}_4 Tr(\Delta^{\dagger}\Delta)_{3_s} Tr(\Delta^{\dagger}\nonumber\\&\quad\,\,\Delta)_{3_a}+\lambda^{\Delta}_5 Tr(\Delta^{\dagger}\Delta)_{3_a} Tr(\Delta^{\dagger}\Delta)_{3_a},\nonumber\\
V(\eta)&=  -\mu^2_{\eta}(\eta^{\dagger}\eta)+ \lambda^{\eta}(\eta^{\dagger}\eta)^2,\nonumber\\
V(\kappa)&=  -\mu^2_{\kappa}(\kappa^{\dagger}\kappa)+ \lambda^{\kappa}(\kappa^{\dagger}\kappa)^2,\nonumber\\
V(\xi)&=  -\mu^2_{\xi}(\xi^{\dagger}\xi)+ \lambda^{\xi}(\xi^{\dagger}\xi)^2,\nonumber\\
V(\zeta)&=  -\mu^2_{\zeta}(\zeta^{\dagger}\zeta)+ \lambda^{\zeta}(\zeta^{\dagger}\zeta)^2,\nonumber\\
V(\phi,\Phi)&=\lambda^{\phi \Phi}_1(\phi^{\dagger}\phi)(\Phi^{\dagger}\Phi)+\lambda^{\phi \Phi}_2((\phi^{\dagger}\phi)_{1'}(\Phi^{\dagger}\Phi)_{1''}+\nonumber\\&\quad\,(\phi^{\dagger}\phi)_{1''}(\Phi^{\dagger}\Phi)_{1'})+\lambda^{\phi \Phi}_3(\phi^{\dagger}\phi)_{3_s}(\Phi^{\dagger}\Phi)_{3_s}+\nonumber\\&\quad\,\lambda^{\phi \Phi}_4((\phi^{\dagger}\phi)_{3_s}(\Phi^{\dagger}\Phi)_{3_a}+(\phi^{\dagger}\phi)_{3_a}(\Phi^{\dagger}\Phi)_{3_s})\nonumber\\&\quad\,+\lambda^{\phi \Phi}_5(\phi^{\dagger}\phi)_{3_a}(\Phi^{\dagger}\Phi)_{3_a},\nonumber\\
V(\phi, \Delta)&= V(\phi, \Phi \rightarrow \phi,\Delta)+(\mu_1 \phi^{T} i \sigma_2 \Delta^{T} \phi+h.c),\nonumber\\
V(\phi,\eta)&=\lambda^{\phi \eta}(\phi^{\dagger}\phi)\eta^{\dagger}\eta,\nonumber\\
 V(\phi,\kappa)&=\lambda^{\phi \kappa}(\phi^{\dagger}\phi)\kappa^{\dagger}\kappa,\nonumber\\
V(\phi,\xi)&=\lambda^{\phi \xi}(\phi^{\dagger}\phi)\xi^{\dagger}\xi,\nonumber\\
 V(\phi,\zeta)&=\lambda^{\phi \zeta}(\phi^{\dagger}\phi)\zeta^{\dagger}\zeta,\nonumber\\
V(\Phi,\Delta)&=V(\phi,\Phi \rightarrow \Phi,\Delta)+(\mu_2 \Phi^{T} i \sigma_2 \Delta^{\dagger} \Phi+h.c),\nonumber\\
V(\Phi,\eta)&=\lambda^{\Phi \eta}(\Phi^{\dagger}\Phi)\eta^{\dagger}\eta,\nonumber\\
V(\Phi,\kappa)&=\lambda^{\Phi \kappa}(\Phi^{\dagger}\Phi)\kappa^{\dagger}\kappa,\nonumber\\
V(\Phi,\xi)&=\lambda^{\Phi \xi}(\Phi^{\dagger}\Phi)\xi^{\dagger}\xi,\nonumber\\
V(\Phi,\zeta)&=\lambda^{\Phi \zeta}(\Phi^{\dagger}\Phi)\zeta^{\dagger}\zeta,\nonumber\\
V(\Delta,\eta)&=\lambda^{\eta \Delta }Tr(\Delta^{\dagger}\Delta)\eta^{\dagger}\eta,\nonumber\\
V(\Delta,\kappa)&=\lambda^{\kappa \Delta}Tr(\Delta^{\dagger}\Delta)\kappa^{\dagger}\kappa,\nonumber\\
V(\Delta,\xi)&=\lambda^{\Delta \xi}Tr(\Delta^{\dagger}\Delta)\xi^{\dagger}\xi,\nonumber\\
V(\Delta,\zeta)&=\lambda^{\Delta \zeta}Tr(\Delta^{\dagger}\Delta)\zeta^{\dagger}\zeta,\nonumber\\
V(\eta,\kappa)&=\lambda^{\eta \kappa}(\eta^{\dagger}\eta \kappa^{\dagger}\kappa +\mu^2_{\eta\kappa}(\eta \kappa+h.c)+(\eta^2 \kappa^2+\nonumber\\&\quad\,h.c)),\nonumber\\
V(\eta,\xi)&=\lambda^{\eta\xi}\eta^{\dagger}\eta \xi^{\dagger}\xi,\quad V(\eta,\zeta)=\lambda^{\eta\zeta}\eta^{\dagger}\eta \zeta^{\dagger}\zeta,\nonumber\\
V(\xi,\zeta)&=V(\eta, \kappa \rightarrow \xi, \zeta),\nonumber\\
V(\kappa,\xi)&=\lambda^{\kappa\xi}\kappa^{\dagger}\kappa \xi^{\dagger}\xi,\quad V(\kappa,\zeta)=\lambda^{\kappa\zeta}\kappa^{\dagger}\kappa \zeta^{\dagger}\zeta,\nonumber\\
V(\phi,\eta,\kappa)&=\lambda^{\phi \eta \kappa}((\phi^{\dagger}\phi)\eta \kappa+h.c),\nonumber\\
V(\phi,\xi,\zeta)&=\lambda^{\phi \xi \zeta}((\phi^{\dagger}\phi)\xi \zeta+h.c),\nonumber\\
V(\Phi,\eta,\kappa)&=\lambda^{\Phi \eta \kappa}((\Phi^{\dagger}\Phi)\eta \kappa+h.c),\nonumber\\
V(\Phi,\xi,\zeta)&=\lambda^{\Phi \xi \zeta}((\Phi^{\dagger}\Phi)\xi \zeta+h.c),\nonumber\\
V(\Delta,\eta,\kappa)&=\lambda^{\Delta \eta \kappa}(Tr(\Delta^{\dagger}\Delta)\eta \kappa+h.c),\nonumber\\
V(\Delta,\xi,\zeta)&=\lambda^{\Delta \xi \zeta}(Tr(\Delta^{\dagger}\Delta)\xi \zeta+h.c),\nonumber\\
V(\eta,\kappa,\xi,\zeta)&=\lambda^{\eta \kappa \xi \zeta}(\eta \kappa \xi \zeta+h.c).
\end{align}
In models incorporating discrete symmetry, it is common to encounter multiple coupling constants in the scalar potential. This allows for the flexibility to choose appropriate vacuum alignments for the scalar fields. Without loss of generality, for the chosen VEVs, $\langle \phi \rangle = v(1,0,0)$, $\langle \Phi\rangle = u(0,1,1)$, $\langle \Delta \rangle= w (0,1,-1)$, $\langle \eta \rangle  =\,v_{\eta}$, $\langle \kappa\rangle \,=\, v_{\kappa}$, $\langle \xi \rangle  =\,v_{\xi}$ and $ \langle \zeta\rangle \,=\, v_{\zeta}$, the following equations are valid from the minimisation conditions of the scalar potential as shown below,
\begin{align}
&\frac{\partial V}{\partial \phi_{1}}= 2 v v_{\eta} v_{\kappa} \lambda ^{\phi \eta \kappa }+v v_{\eta}^2 \lambda ^{\phi \eta }+v v_{\kappa}^2 \lambda ^{\phi \kappa }+2 u^2 v \lambda _1^{\phi \Phi }-\nonumber\\&\quad\quad\quad \frac{4}{9}u^2 v \lambda _3^{\phi \Phi }+2 v^3 \lambda _1^{\phi }+\frac{8}{9} v^3 \lambda _3^{\phi }-v \mu _{\phi }^2-2 v w^2 \lambda _1^{\phi \Delta }\nonumber\\&\quad\quad\quad +\frac{4}{9} v w^2 \lambda _3^{\phi \Delta }+v \lambda ^{\phi \xi } v_{\xi}^2+2 v \lambda ^{\phi \xi \zeta } v_{\xi} v_{\zeta}+v \lambda ^{\phi \zeta }\nonumber\\&\quad\quad\quad  v_{\zeta}^2=0,\\
&\frac{\partial V}{\partial \phi_{2}}=u^2 v \lambda _2^{\phi \Phi }-\frac{2}{9} u^2 v \lambda _3^{\phi \Phi }+\frac{1}{3} u^2 v \lambda _4^{\phi \Phi }+v w^2 \lambda _2^{\phi \Delta }-\frac{2}{9}\nonumber\\&\quad\quad\quad v w^2 \lambda _3^{\phi \Delta }+\frac{1}{3} v w^2 \lambda _4^{\phi \Delta }-\frac{4}{3} \mu _1 v w=0,\\
&\frac{\partial V}{\partial \phi_{3}}=u^2 v \lambda _2^{\phi \Phi }-\frac{2}{9} u^2 v \lambda _3^{\phi \Phi }-\frac{1}{3} u^2 v \lambda _4^{\phi \Phi }+v w^2 \lambda _2^{\phi \Delta }-\frac{2}{9}\nonumber\\&\quad\quad\quad v w^2 \lambda _3^{\phi \Delta }-\frac{1}{3} v w^2 \lambda _4^{\phi \Delta }+\frac{4}{3} \mu _1 v w=0,\\
&\frac{\partial V}{\partial \Phi_{1}}=2 u^3 \lambda _2^{\Phi }-\frac{8}{9} u^3 \lambda _3^{\Phi }+2 u w^2 \lambda _2^{\Phi \Delta }-\frac{4}{9} u w^2 \lambda _3^{\Phi \Delta }=0,\\
&\frac{\partial V}{\partial \Phi_{2}}=2 u v_{\eta} v_{\kappa} \lambda^{\Phi \eta \kappa }+u v_{\eta}^2 \lambda^{\Phi \eta }+u v_{\kappa}^2 \lambda ^{\Phi \kappa }+4 u^3 \lambda _1^{\Phi }+u^3 \nonumber\\&\quad\quad\quad\lambda _2^{\Phi }+\frac{4}{3} u^3 \lambda _3^{\Phi }+\frac{1}{3} u^3 \lambda _4^{\Phi }-u \mu _{\Phi }^2+u v^2 \lambda _1^{\phi \Phi }-\frac{2}{9} u v^2\nonumber\\&\quad\quad\quad \lambda _3^{\phi \Phi }-\frac{1}{3} u v^2 \lambda _4^{\phi \Phi }+u \lambda ^{\Phi \xi } v_{\xi}^2+2 u \lambda ^{\Phi \xi \zeta } v_{\xi} v_{\zeta}+u \lambda ^{\Phi \zeta }\nonumber\\&\quad\quad\quad v_{\zeta}^2-2 u w^2 \lambda _1^{\Phi \Delta }+u w^2 \lambda _2^{\Phi \Delta }+\frac{2}{9} u w^2 \lambda _3^{\Phi \Delta }-\frac{1}{3} u w^2\nonumber\\&\quad\quad\quad \lambda _4^{\Phi \Delta }-\frac{8}{3} \mu _2 u w=0,\\
&\frac{\partial V}{\partial \Phi_{3}}=2 u v_{\eta} v_{\kappa}\lambda ^{\Phi \eta \kappa }+u v_{\eta}^2 \lambda ^{\Phi \eta }+u v_{\kappa}^2 \lambda ^{\Phi \kappa }+4 u^3 \lambda _1^{\Phi }+u^3\nonumber\\&\quad\quad\quad \lambda _2^{\Phi }+\frac{4}{3} u^3 \lambda _3^{\Phi }-\frac{1}{3} u^3 \lambda _4^{\Phi }-u \mu _{\Phi }^2+u v^2 \lambda _1^{\phi \Phi }-\frac{2}{9} u v^2 \nonumber\\&\quad\quad\quad \lambda _3^{\phi \Phi }+\frac{1}{3} u v^2 \lambda _4^{\phi \Phi }+u \lambda ^{\Phi \xi } v_{\xi}^2+2 u \lambda ^{\Phi \xi \zeta } v_{\xi} v_{\zeta}+u \lambda ^{\Phi \zeta }\nonumber\\&\quad\quad\quad v_{\zeta}^2-2 u w^2 \lambda _1^{\Phi \Delta }+u w^2 \lambda _2^{\Phi \Delta }+\frac{2}{9} u w^2 \lambda _3^{\Phi \Delta }+\frac{1}{3} u w^2 \nonumber\\&\quad\quad\quad\lambda _4^{\Phi \Delta }+\frac{8}{3} \mu _2 u w=0,\\
&\frac{\partial V}{\partial \Delta_{1}}=\frac{-2}{3} \left(-2 \mu _2 u^2+u^2 w \lambda _4^{\Phi \Delta }+2 \mu _1 v^2+w^3 \lambda _4^{\Delta }\right)=0,\nonumber\\
&\frac{\partial V}{\partial \Delta_{2}}=-2 w \lambda ^{\Delta \eta \kappa } v_{\eta} v_{\kappa}-w v_{\eta}^2 \lambda ^{\eta \Delta }-w \lambda ^{\kappa \Delta } v_{\kappa}^2-\frac{4 \mu _2 u^2}{3}-\nonumber\\&\quad\quad\quad 2 u^2 w \lambda _1^{\Phi \Delta }+u^2 w \lambda _2^{\Phi \Delta }+\frac{2}{9} u^2 w \lambda _3^{\Phi \Delta }-\frac{1}{3} u^2 w \lambda _4^{\Phi \Delta }-\nonumber\\&\quad\quad\quad v^2 w \lambda _1^{\phi \Delta }+\frac{2}{9} v^2 w \lambda _3^{\phi \Delta }+\frac{1}{3} v^2 w \lambda _4^{\phi \Delta }-w \lambda ^{\Delta \xi } v_{\xi}^2-2 \nonumber\\&\quad\quad\quad w \lambda ^{\Delta \xi \zeta } v_{\xi} v_{\zeta}-w \lambda ^{\Delta \zeta } v_{\zeta}^2+4 w^3 \lambda _1^{\Delta }+w^3 \lambda _2^{\Delta }+\frac{4}{3} w^3 \nonumber\\&\quad\quad\quad \lambda _3^{\Delta }+\frac{1}{3} w^3 \lambda _4^{\Delta }+w \mu _{\Delta }^2=0,
\end{align}

\begin{align}
&\frac{\partial V}{\partial \Delta_{3}}=2 w \lambda ^{\Delta \eta \kappa } v_{\eta} v_{\kappa}+w v_{\eta}^2 \lambda ^{\eta \Delta }+w \lambda ^{\kappa \Delta } v_{\kappa}^2-\frac{4 \mu _2 u^2}{3}+2\nonumber\\&\quad\quad\quad u^2 w \lambda _1^{\Phi \Delta }-u^2 w \lambda _2^{\Phi \Delta }-\frac{2}{9} u^2 w \lambda _3^{\Phi \Delta }-\frac{1}{3} u^2 w \lambda _4^{\Phi \Delta }+\nonumber\\&\quad\quad\quad v^2 w \lambda _1^{\phi \Delta }-\frac{2}{9} v^2 w \lambda _3^{\phi \Delta }+\frac{1}{3} v^2 w \lambda _4^{\phi \Delta }+w \lambda ^{\Delta \xi } v_{\xi}^2+2 \nonumber\\&\quad\quad\quad w \lambda ^{\Delta \xi \zeta } v_{\xi} v_{\zeta}+w \lambda ^{\Delta \zeta } v_{\zeta}^2-4 w^3 \lambda _1^{\Delta }-w^3 \lambda _2^{\Delta }-\frac{4}{3} w^3 \nonumber\\&\quad\quad\quad \lambda _3^{\Delta }+\frac{1}{3} w^3 \lambda _4^{\Delta }-w \mu _{\Delta }^2=0,\\
&\frac{\partial V}{\partial \eta}=2 v_{\eta}(-\mu _{\eta}^2+3 \lambda ^{\eta \kappa } v_{\kappa}^2+2 u^2 \lambda ^{\Phi \eta }+v^2 \lambda ^{\phi \eta }+\lambda ^{\eta \xi } v_{\xi}^2+ \nonumber\\&\quad\quad\quad\lambda ^{\eta \zeta }v_{\zeta}^2-2 w^2 \lambda ^{\eta \Delta })+4 \lambda ^{\eta} v_{\eta}^3-2 v_{\kappa}(-\lambda ^{\eta \kappa } \mu _{\eta \kappa }^2-2\nonumber\\&\quad\quad\quad u^2 \lambda ^{\Phi \eta \kappa }-v^2 \lambda ^{\phi \eta \kappa }-\lambda ^{\eta \kappa \xi \zeta } v_{\xi} v_{\zeta}+2 w^2 \lambda ^{\Delta \eta \kappa })=0,\\
&\frac{\partial V}{\partial \kappa}=2(3 v_{\eta}^2 \lambda ^{\eta \kappa } v_{\kappa}+v_{\eta}(\lambda ^{\eta \kappa } \mu _{\eta \kappa }^2+2 u^2 \lambda ^{\Phi \eta \kappa }+v^2 \lambda ^{\phi \eta \kappa }+\nonumber\\&\quad\quad\quad\lambda ^{\eta \kappa \xi \zeta } v_{\xi} v_{\zeta}-2 w^2 \lambda ^{\Delta \eta \kappa })+v_{\kappa}(-\mu _{\kappa}^2+2 \lambda ^{\kappa} v_{\kappa}^2+2 u^2\nonumber\\&\quad\quad\quad \lambda ^{\Phi \kappa }+v^2 \lambda ^{\phi \kappa }+\lambda ^{\kappa \xi } v_{\xi}^2+\lambda ^{\kappa \zeta } v_{\zeta}^2-2 w^2 \lambda ^{\kappa \Delta }))=0,\\
&\frac{\partial V}{\partial \xi}=2 v_{\xi}(v_{\eta}^2 \lambda ^{\eta \xi }+\lambda ^{\kappa \xi } v_{\kappa}^2+2 u^2 \lambda ^{\Phi \xi }+v^2 \lambda ^{\phi \xi }+3 \lambda ^{\xi \zeta } v_{\zeta}^2-\nonumber\\&\quad\quad\quad2 w^2 \lambda ^{\Delta \xi }-\mu _{\xi}^2)-2 v_{\zeta}(-v_{\eta} \lambda ^{\eta \kappa \xi \zeta } v_{\kappa}-\lambda ^{\xi \zeta } \mu _{\xi \zeta }^2-2 u^2\nonumber\\&\quad\quad\quad \lambda ^{\Phi \xi \zeta }-v^2 \lambda ^{\phi \xi \zeta }+2 w^2 \lambda ^{\Delta \xi \zeta })+4 v_{\xi}^3 \lambda ^{\xi}=0,\\
&\frac{\partial V}{\partial \zeta}=2(v_{\xi}(v_{\eta} \lambda ^{\eta \kappa \xi \zeta } v_{\kappa}+\lambda ^{\xi \zeta } \mu _{\xi \zeta }^2+2 u^2 \lambda ^{\Phi \xi \zeta }+v^2 \lambda ^{\phi \xi \zeta }-\nonumber\\&\quad\quad\quad2 w^2 \lambda ^{\Delta \xi \zeta })+v_{\zeta}(v_{\eta}^2 \lambda ^{\eta \zeta }+\lambda ^{\kappa \zeta } v_{\kappa}^2+2 u^2 \lambda ^{\Phi \zeta }+v^2 \lambda ^{\phi \zeta }\nonumber\\&\quad\quad\quad+2 v_{\zeta}^2 \lambda ^{\zeta}-2 w^2\lambda ^{\Delta \zeta }-\mu _{\zeta}^2)+3 \lambda ^{\xi \zeta } v_{\xi}^2 v_{\zeta})=0.
\end{align}
\end{document}